\documentclass[aps,preprint,amssymb]{revtex4}

\usepackage{epsfig}
\usepackage{graphicx}

\begin{document}
\renewcommand{\theequation}{\arabic{section}.\arabic{equation}}
\newcommand{\be}{\begin{equation}}
\newcommand{\ee}{\end{equation}}
\newcommand{\bea}{\begin{eqnarray}}
\newcommand{\eea}{\end{eqnarray}}
\newcommand{\bi}{\bibitem}
\newcommand{\ddg}{\ddagger}

\renewcommand{\r}{({\bf r})}
\newcommand{\rp}{({\bf r'})}

\newcommand{\ua}{\uparrow}
\newcommand{\da}{\downarrow}
\newcommand{\la}{\langle}
\newcommand{\ra}{\rangle}
\newcommand{\dg}{\dagger}

\title{Excitation energies from ground-state density-functionals by means of generator coordinates}

\author{E. Orestes,$^{a,b}$ A. B. F. da Silva$^a$ and K. Capelle$^{b}$}
\email[]{E-mail: capelle@ifsc.usp.br}

\affiliation{$^a$ Departamento de F\'{\i}sica e Qu\'{\i}mica, Instituto de Qu\'{\i}mica de S\~ao Carlos, Universidade de S\~ao Paulo, Caixa Postal 780, 13560-970 S\~ao Carlos, SP, Brazil\\
$^b$ Departamento de  F\'{\i}sica e Inform\'atica, Instituto de F\'{\i}sica de S\~ao Carlos, Universidade de S\~ao Paulo, Caixa Postal 369, 13560-970 S\~ao Carlos, SP, Brazil}

\date{\today}

\begin{abstract}
The generator-coordinate method is a flexible and powerful reformulation of the variational principle. Here we show that by introducing a generator coordinate in the Kohn-Sham equation of density-functional theory, excitation energies can be obtained from ground-state density functionals. As a viability test, the method is applied to ground-state energies and various types of excited-state energies of atoms and ions from the He and the Li isoelectronic series. Results are compared to a variety of alternative DFT-based approaches to excited states, in particular time-dependent density-functional theory with exact and approximate potentials.
\end{abstract}

\maketitle

\section{Introduction}
\label{intro}

Density-functional theory (DFT) \cite{kohnrmp,dftbook,parryang} is the most widely used first-principles method for the calculation of ground-state properties of a wide variety of atomic, molecular and solid systems. Common local and nonlocal density functionals make highly precise predictions for ground-state densities, energies and quantities that can be derived from these. Efficient algorithms for solving the Kohn-Sham (KS) equations are implemented in many common electronic-structure codes, and provide computational access to ground-state densities by means of an auxiliary set of single-particle orbitals.

The Kohn-Sham equations, as well as most common density functionals, were designed with ground-state properties in mind. The underlying theorems of density-functional theory, however, are more general. The usual proofs of the Hohenberg-Kohn theorem by contradiction \cite{hk} or by constrained search \cite{levylieb} guarantee that the ground-state wave function is a functional of the ground-state density, but it takes only one simple additional step to prove that the density also determines the external potential \cite{dftbook,levy,nonunprl} and thus all wave functions and energies, including those of excited states \cite{footnote01}.

DFT thus holds the promise to become a versatile and powerful tool for the calculation of excited-state energies, too. In practice, however, this promise turned out to be much harder to fulfill than for the ground state. In fact, the most commonly employed DFT approaches to excited states are formulated in the conceptually rather different frameworks of time-dependent DFT \cite{tddftex} or ensemble DFT \cite{ogk}. For a comparison of these methods, and a variety of other DFT-based approaches to excited states, see Ref.~\cite{otherex}. What TDDFT and ensemble DFT have in common is that by design they go, from the outset, beyond static (ground-state) DFT. In this paper we propose and test an alternative method that allows one to extract excited-state energies from ordinary ground-state functionals.

\section{The generator coordinate variational principle}
\label{gcmdft}

The key ingredient in this development is the generator coordinate variational principle. The generator-coordinate method (GCM) arose in nuclear physics, as a way to build collective behaviour of nuclei into a trial wave function written in terms of single-particle orbitals \cite{ghw}. In this method, the (nuclear) many-body wave function $\Psi$ is cast as
\be
\Psi(x_1,..,x_N)=
\int d\alpha\, f(\alpha)\Phi(\alpha;x_1,..,x_N),
\label{ghwansatz1}
\ee
where the $\Phi$ are auxiliary wave functions arising from a deformed Hamiltonian, and the degree of deformation is characterized by the deformation parameter $\alpha$, which plays the role of a generator coordinate. It is a crucial feature of the method that the generator coordinate appears in the auxiliary (seed) functions $\Phi$, but not in the wave function $\Psi$. Variation of the energy with respect to the weight function $f(\alpha)$ leads to an integral equation whose eigenvalues are the many-body energies of the system,
\be
\int d\alpha'\,\left[K(\alpha,\alpha')-ES(\alpha,\alpha') \right]
f(\alpha') = 0.
\label{ghweq}
\ee
Equation (\ref{ghweq}), known as the Griffin-Hill-Wheeler (GHW) equation, contains Hamiltonian and overlap kernels $K(\alpha,\alpha')= \la \Phi(\alpha)| \hat{H} | \Phi(\alpha') \ra$ and $S(\alpha,\alpha')=\la \Phi(\alpha)| \Phi(\alpha') \ra$ of standard form, and can be solved straightforwardly by discretizing the $\alpha$ integral and obtaining the eigenvalues $E$ and eigenfunctions $f(\alpha)$ by matrix algebra.

The GHW method is also used in quantum chemistry to construct highly precise basis sets for Hartree-Fock and Dirac-Fock calculations \cite{ghwhf}. In these
applications the {\em single-particle} orbitals are written in GCM form as
\be
\varphi(x)=
\int d\alpha\, f(\alpha)\xi(\alpha;x),
\label{ghwansatz2}
\ee
where $\xi(\alpha)$ is a set of suitable single-body functions, $\alpha$ is identified with a basis set exponent, and $\varphi(x)$ is the Hartree-Fock (or Dirac-Fock) orbital. The resulting GHW equation produces numerically defined basis sets of high accuracy \cite{ghwhf}.

In 2003 one of the present authors proposed a way to combine the GCM variational principle with DFT \cite{gencoojcp}, by identifying the deformation potential of the original GCM with the KS potential of DFT, the generating function with the KS Slater determinant, and the generator coordinate with any parameter in the effective single-particle potential. Hence,
\be
\Psi(x_1,..,x_N)=
\int d\alpha\, f(\alpha)\Phi(\alpha;x_1,..,x_N),
\label{ghwansatz3}
\ee
where $\Psi$ now is the electronic many-body wave function, and $\Phi$ is the KS Slater determinant.  Conceptually, this is much more similar to the original use of GCM in nuclear physics \cite{ghw} than to its more recent applications in quantum chemistry \cite{ghwhf}.

A first implementation of GCM-DFT \cite{gencoojcp} showed that even simple approximations to the full formalism result in ground-state energies that are comparable to those obtained from more sophisticated density functionals. Moreover, unlike standard DFT methods, GCM-DFT also provides an approximation to the many-body wave function, not just to the energies. The final expressions obtained for these wave functions are formally similar to CI expansions, but whereas in CI each determinant represents a different excited-state configuration of one fixed Hamiltonian, in the GCM-DFT method of Ref.~\cite{gencoojcp} each determinant comes from a ground-state calculation of a differently deformed Hamiltonian, and can thus be viewed as a resummation of many CI determinants. As these GCM determinants come from different Hamiltonians they are not necessarily orthogonal, in contrast with CI determinants. While at first sight this increases the computational effort involved in constructing the GCM-DFT expansion relative to that for constructing the CI expansion, the fact that each GCM determinant can be viewed as a resummation of many CI determinants suggests that much smaller expansions may be sufficient with GCM than with CI, in particularly if the seed functions are cleverly chosen.

Ref.~\cite{gencoojcp} also showed that approximations to the energies of excited states could be obtained as by-products of the ground-state calculation, but the numerical example given there showed that this method performs rather poorly. Here we present a reformulation of GCM-DFT which performs much better for excited states -- in fact, even in its simplest form it is already competitive with standard TDDFT based methods.

\section{GCM-DFT for excited states}
\label{gcmdftex}

To introduce the key idea, let us first briefly review the way ground-state energies are obtained from GCM-DFT \cite{gencoojcp}. The seed wave functions (or generator functions) $\Phi(\alpha;x_1,..,x_N)$ are chosen to be the Slater determinants arising from Kohn-Sham orbitals of the N lowest-lying levels of the deformed KS Hamiltonian, i.e., are KS $N$-particle ground states. The particular deformation chosen in Ref.~\cite{gencoojcp} was
\be
v_\alpha\r = v_{ext}\r + v_H\r + \alpha v_x^{LDA}\r,
\label{valpha0}
\ee
{\em i.e.}, the exchange-only LDA was modulated by $\alpha$, for which the five-point mesh $\{0,0.5,1,1.5,2\}$ was used. We stress that this is not a $X\alpha$
calculation, although the last term in Eq.~(\ref{valpha0}) is of $X\alpha$ form: $\alpha$ here is a generator coordinate, which appears as integration variable in the integral-representation of the wave function (\ref{ghwansatz3}).
Upon discretization of the integral, $\alpha$ assumes more than one value,
none of which is fitted to experiment. For each $\alpha$ we selfconsistently
solve the KS equations, construct $\Phi(\alpha)$ from the resulting orbitals
and calculate the kernels $K(\alpha,\alpha')$ and
$S(\alpha,\alpha')$.

The benchmark He ground-state energy $E_0^{He}=-2.904a.u.$ was reproduced in this way to within $1.1\%$. A further improvement on the ground-state energy, reducing the deviation to $-0.24\%$, is reported below. (For comparison, the LDA, when used in the standard DFT way without generator coordinates, predicts $E_0^{He,LDA}=-2.8348 a.u.$, which deviates by -2.4\% from $E_0^{He}$.)

In principle, the five eigenvalues arising from the $5\times 5$ matrix problem allow one to obtain four excited-state energies, in addition to the ground state, but the lowest-lying excitation energy was found \cite{gencoojcp} to deviate by $16.7\%$ from the exact value, which is too much to be useful in practice.

However, we note that the trial function $\Psi$ inherits its symmetries and structure from the seed functions $\Phi(\alpha)$. We can thus target a particular excited state of the many-body system by using seed functions of the corresponding excited KS state, and solving the GHW equation with kernels obtained from these excited-state determinants or configuration-state functions. This procedure can be used for all many-body excitations that have a counterpart in the noninteracting KS system. (Excitations not having such a counterpart can be dealt with by using a generator coordinate in TDDFT \cite{tdgcm}, or by using more sophisticated seed functions in static GCM-DFT.)

Determinants corresponding to excited states of the noninteracting KS system are rarely used in DFT, because the KS formalism is tailored to provide ground-state properties. However, in the context of GCM-DFT, the KS equation is not used to obtain the ground-state density, but to obtain a set of continuously parametrized $N$-particle determinants that are employed as seed functions $\Phi(\alpha)$ in the GCM {\em ansatz} Eq.~(\ref{ghwansatz1}), and for this purpose the use of excited-state KS determinants is perfectly legitimate.

\section{Applications to representative two- and three-electron systems}
\label{applications}

In this section we report numerical results for different types of excitation
energies (singlet/doublet/triplet) as well as ground-state energies, for
various two- and three-electron systems from the He and Li isoelectronic series,
compared to a variety of other computational approaches, including TDDFT.

\subsection{Excited-state energies for atoms from the He isoelectronic series}
\label{heappl}

First, we present illustrative applications to excited states of the Helium isoelectronic series, and to the lowest singlet and triplet excitations of the He atom. We chose the deformation potential to be
\be
v_\alpha\r = v_{ext}\r + v_H\r + \alpha [v_x^{LDA}\r+v_c^{LDA}\r],
\label{valpha}
\ee
{\em i.e.}, let the generator coordinate modulate the LDA for exchange and correlation. Many other choices are possible, {\em e.g.}, introducing $\alpha$ in the correlation potential, Hartree potential, external potential, kinetic energy term, or in the angular or spin-dependent part of the orbitals, or using other functionals than LDA. Each of the resulting deformations has its own physical meaning and consequences. Each also implies a, generally distinct, range of values of $\alpha$, and, consequently, different meshes for discretizing the integral equation.

Some choices of where to place $\alpha$ in the KS equations, and some discretization schemes for the $\alpha$ integral, give excellent results for specific quantities or some particular system. Below we do not report such special choices or best results, but focuse on a simple and generally usable scheme \cite{footnote02}. Systematic exploration of the many possibilities that arise upon combining the GCM idea with DFT remains work for the future.

In Table~\ref{table1}, we report lowest triplet excited-state energies of the Helium isoelectronic series, obtained from solving the discretized GHW eigenvalue equation with $v_\alpha\r$ chosen as in Eq.~(\ref{valpha}), on the mesh $\{4.5,5,5.5,6,6.5\}$. (Denser meshes lead to only marginal improvements of the results, or may even worsen them if the resulting overlap matrix $S(\alpha,\alpha')$ becomes singular.) The seed wave functions were constructed from the KS orbitals arising selfconsistently in potential (\ref{valpha}), selected to form the lowest KS excited state of triplet symmetry. Near-exact theoretical data are also reported, and show that the resulting excited-state energies are surprisingly close. Table~\ref{table2} makes the same comparison for the lowest singlet excitation, $2^1S$.

For spectroscopy, the interesting quantities are not primarily the excited-state energies, but the excitation energies. In Table~\ref{table3} we thus present selected excitation energies of the He atom: the lowest singlet and triplet excitation energies, and the singlet-triplet splitting.
GCM-DFT can be used as a {\em stand-alone} method to obtain excitation energies if it is used to calculate the energies of the ground state and of the excited state of interest. Alternatively, it can be used as an {\em add-on} method by adding excitation energies obtained from GCM-DFT to ground-state energies obtained by traditional methods. Both procedures are compared in Table~\ref{table3}
to six other computational schemes and near-exact benchmark data.

The columns KS exact and KS LDA report the KS single-particle gap between the highest occupied and lowest unoccupied KS eigenvalue, arising from the exact \cite{pgg} KS potential and from the LDA potential for He, respectively. Note that neither the LDA nor the exact KS eigenvalues predict any singlet-triplet splitting.
The column labeled $\Delta$SCF reports LDA total-energy differences between the ground state and the total energy obtained from a standard KS calculation fixing the occupation of KS levels at that of the corresponding KS excited state.

The two columns labelled TDDFT**EXX and TDDFT**ALDA report data obtained in Ref.~\cite{pgg} from TDDFT using adiabatic exact-exchange (AEXX) and adiabatic LDA (ALDA), respectively. These TDDFT data
were obtained from a numerically exact potential \cite{pgg}, approximating only the xc kernel, and in this sense are not fully representative of standard TDDFT calculations. ALDA results obtained from the standard TDDFT procedure, using approximate potentials and kernels, are reported in the column labelled TDDFT ALDA.
Our LDA and TDDFT ALDA data were obtained with the {\tt GAUSSIAN 03} \cite{gaussian} program, using the VWN5 parametrization of the LDA and
the $aug-ccpV5Z$ basis set, and with the mesh-based atomic DFT code {\em opmks} \cite{opmks}.

The column labeled GCM-DFT* reports differences between GCM-DFT excited-state energies and the exact ground-state energy, whereas the column labelled GCM-DFT reports differences between GCM-DFT excited-state energies and GCM-DFT ground-state energies, the latter being obtained from Eq.~(\ref{valpha}) on the mesh $\{4.7,5.05,5.4,5.75,6.1\}$. The He ground-state energy obtained on this mesh is $-2.897a.u.$, which deviates from the exact result $-2.904a.u.$ by $-0.24\%$.
The column labeled GCM-DFT* thus measures the {\em add-on} performance of generator-coordinate DFT, in our simple implementation, purely for excited states, while the column labeled GCM-DFT quantifies its {\em stand-alone} performance for excitation energies.

As the data in the lines labelled $^3S$ and $^2S$ of Table~\ref{table3} show,
GCM-DFT, both in its {\em add-on} and its {\em stand-alone} version, is capable
of producing similar or better singlet and triplet excited-state
energies than the other tested methods. As the line labelled $\Delta$ shows,
it also produces realistic
singlet-triplet splittings. Still better singlet-triplet splittings are
obtained from TDDFT/ALDA, both with approximate and exact single-particle
potentials, but this improvement is due to error cancellation between the
energy of the singlet and the triplet excited state, each of which individually
has larger errors.

Note that we have not separately optimized the mesh for singlet excitations and triplet excitations, but used the same five values of $\alpha$ for both. In other calculations, we have obtained better excitation energies by employing different meshes for different states, but our purpose here is to keep the procedure as simple and generally applicable as possible, and we thus used the same set of $\alpha$'s for all excited states of He.

\subsection{Ground-state and excitation energies for atoms from the Li
isoelectronic series}
\label{liappl}

As an initial exploration of three-electron systems, we now present GCM-DFT
calculations of the ground-state energy and excitation energies of atoms from the Li isoelectronic
series. In principle, one could choose a new placement of the generator coordinate in the KS equation, and a new mesh, for each new system. In this initial exploration, we maintained for Li the choices made for He. Ground-state energies obtained in this way are reported in the first column of Table~\ref{table4}, labelled `He mesh'. Clearly, the deviations from the experimental energies are much larger than they were for the He series.

Next, in recognition of the fact that the average radius $\la r \ra$ of an
atom shrinks as $Z$ increases (for one-electron atoms, $\la r \ra \propto 1/Z$), and taking into account that the generator coordinate in Eq.~(\ref{valpha})
modifies only the radial wave function, we constructed downscaled radial meshes for Li by applying a simple power-law scaling factor to the mesh for He,
\be
\alpha(Li) = \alpha(He)\left(\frac{Z_{He}}{Z_{Li}}\right)^{\frac{3}{4}},
\label{HeLigsLaw1}
\ee
where the exponent $3/4$ was chosen after some experimentation (but not
optimized variationally). This scaling approach is quite successful: The
second set of data in Table~\ref{table4} has much reduced deviations from
the benchmark data.

Finally, instead of adjusting the exponent of the mesh scaling law
(\ref{HeLigsLaw1}), we also constructed a new discretization mesh for
neutral Li, which was subsequently used also for the ions.
The free parameters here are the starting value and the increment, {\em i.e.}
one more than in the scaling approach. Results are shown in the third
data set in Table~\ref{table4}, labelled `emp', for empirical (but not
fitted). As expected, the agreement with
benchmark data is further improved. We stress that in using these modified
meshes we still employed the same generator coordinate as in
Eq.~(\ref{valpha}) and still constrained the wave function to be a sum of
five terms, corresponding to five values of $\alpha$, {\em i.e.}, our choice
of mesh is an exploration of the capabilities of the GCM method, not an
arbitrary construction designed to produce the best possible agreement
with reference data.

Turning now from ground states to excited states, Table~\ref{table5} presents
KS, $\Delta$SCF, TDDFT and GCM-DFT data for
excitation energies corresponding to the ground-state to doublet-excited-state
$^2S$ transition $1s^{2}2s^{1} \rightarrow 1s^{2}3s^{1}$.
As in Table~\ref{table4},
the first group of Li GCM-DFT data represent the {\em stand-alone} approach,
employing the He mesh, a power-law scaled He mesh and a mesh whose initial value
and step size were chosen to deliver values close to the benchmark data for
neutral Li,
but subject to the constraints of same placement of the generator coordinate
and same number of mesh points. The only difference is that for the excited
states the scaling law exponent $1/3$, leading to the mesh
\be
\alpha(Li) = \alpha(He)\left(\frac{Z_{He}}{Z_{Li}}\right)^{\frac{1}{3}},
\label{HeLigsLaw2}
\ee
was found to be more suitable than $3/4$. The second group of GCM-DFT data,
labelled GCM-DFT*, reports corresponding results from the {\em add-on} use
of GCM, in which only the excited-state energy was obtained from GCM-DFT,
the ground-state energy employed was the exact one.

As for He, GCM-DFT produces realistic excitation energies, but the
agreement with the reference values is less good than for He.
Nevertheless, since we have not fitted, and neither variationally optimized,
the mesh of $\alpha$ values, but just explored some simple and generalizable
discretization schemes, the agreement achieved is still rather encouraging.
Moreover, we
note that the smaller errors TDDFT achieved for the $2^2S \to 3^2S$
excitation energies are a consequence of error cancellation between the
ground-state energy and the excited-state energy, each of which is less well
reproduced by TDDFT/ALDA than by the GCM-DFT calculations. This is
reminiscent of the error cancellation that occured between singlet and triplet
excited-state energies for He.

\section{Conclusions}
\label{concl}

Whether GCM-DFT will ever be competitive with more established TDDFT or CI
approaches to excited states is, at present, an open question, but from the present analysis it seems safe to conclude that, as a matter of principle, excitation energies can indeed be obtained from static ground-state density functionals, without requiring time- or temperature-dependent generalizations, as anticipated by the original Hohenberg-Kohn theorem.

In our opinion, what is missing to turn the presently proposed approach into
a viable and transferable method for calculating ground-state and excitation
energies (and other properties \cite{footnote03}) of many-electron
systems is a systematic and nonempirical way of constructing suitable meshes
for discretizing the $\alpha$ integral in the GHW equation. Different ways of
achieving this are conceivable:

(i) Much denser meshes, employing hundreds or more values of $\alpha$, would
lead to an ever more faithful representation of the GHW integral itself, but in
this case much care must be taken in inverting the overlap matrix, because
closely spaced values of $\alpha$ lead to near linear dependence of rows and
columns and thus to potentially ill-conditioned eigenvalue problems.

(ii) Variational optimization of a set of $\alpha$ values may be useful for the
ground-state energy, and produces a mesh that can be used as a starting point
also for excited states.

(iii) Scaling laws, as our present Eqs.~(\ref{HeLigsLaw1}) and
(\ref{HeLigsLaw2}),
provide a way to use chemical or physical intuition to pre-select a suitable
set of values of the generator coordinate  without data-fitting or
optimization.

This discussion shows that the great strength of the GCM approach is
also its Achilles heel: the fact that the generator coordinate does not
explicitly appear in the wave function makes it a particularly powerful and
flexible formulation of the variational principle, but also implies that
there is no direct way to make an {\em a priori} reasonable choice of
placement and range of the generator coordinate.

We end this paper by noting that, very recently, generator coordinates have also been introduced in TDDFT in order to describe retardation effects and memory by means of a time-dependent generalization of the GHW equations \cite{tdgcm}. A further interesting possibility is to introduce a generator coordinate also in the TDDFT approach to excited states in order to obtain the time-dependent many-body wave function, or improved excitation energies.

This work was supported by FAPESP, CAPES and CNPq. We thank C. A. Ullrich for useful discussions.

\clearpage

\begin{table}
\begin{center}
\caption{\label{table1} Energy of the lowest excited triplet state, $2^3S$, of ions from the He isoelectronic series obtained by GCM-DFT from the LDA, compared to available near-exact theoretical data \cite{expdata}.}
\begin{tabular}{r|cccccccc}
\hline
& $-E^{GCM-DFT}$ & $-E^{ex}$ & $\%$ deviation \\ \hline
$He$ & 2.173  & 2.175 & -0.092\,\%\\
$Li^+$ & 5.109 & 5.104 & 0.098\,\% \\
$Be^{2+}$ &9.294  & 9.289 & 0.054\,\%  \\
$B^{3+}$ & 14.73 & 14.72 & 0.068\,\% \\
$C^{4+}$ & 21.42 & 21.41 & 0.047\,\%  \\
\hline
\end{tabular}
\end{center}
\end{table}

\begin{table}
\begin{center}
\caption{\label{table2} Energy of the lowest excited singlet state, $2^1S$, of ions from the He isoelectronic series obtained by GCM-DFT from the LDA, compared to available near-exact theoretical data \cite{expdata}.}
\begin{tabular}{r|cccccccc}
\hline
& $-E^{GCM-DFT}$ & $-E^{ex}$ & $\%$ deviation \\
\hline
$He$ & 2.137  & 2.146 & -0.419 \,\%\\
$Li^+$ & 5.028 & 5.042 & -0.278 \,\% \\
$Be^{2+}$ &9.170  & 9.181 & -0.120 \,\%  \\
$B^{3+}$ & 14.56 & 14.57 & -0.069 \,\% \\
$C^{4+}$ & 21.21 & 21.21 & 0.000\,\%  \\
\hline
\end{tabular}
\end{center}
\end{table}

\begin{table}
\begin{center}
\caption{\label{table3} Lowest triplet and singlet excitation energies and singlet-triplet splittings of the He atom, obtained by eight different calculational approaches, described in the main text, and near-exact benchmark data. Second line for each excitation: percentage deviation from benchmark data \cite{expdata}. The last two rows report the singlet-triplet splitting $\Delta$ (multiplied by 100 for legibility) and its percentage error.}
\begin{tabular}{ccccccccccc}
\hline
       $He$ & KS & KS & $\Delta$SCF & TDDFT** & TDDFT** & TDDFT & GCM-DFT*  & GCM-DFT & exact\\
$1s \to 2s$ & exact & LDA & LDA      & AEXX    & ALDA  & ALDA & LDA      & LDA  & \\
\hline
$^3S$ & 0.7460 & 0.6218 & 0.7146 & 0.7207 & 0.7351 & 0.6100 & 0.7312 & 0.7240 & 0.7285 & \\
\%    & 2.4    & -14.7  & -1.9   & -1.1   & 0.91   & -16    & 0.37   & -0.62  & - \\
\hline
$^1S$ & 0.7460 & 0.6218 & 0.7292 & 0.7659 & 0.7678 & 0.6373 & 0.7667 & 0.7600 & 0.7578 & \\
\%    & -1.6   & -17.9  & -3.8   & 1.1    &  1.3   & -16    & -1.2   & 0.29   & - \\
\hline
100 $\Delta$ & 0 & 0    & 1.46   & 4.52   & 3.27   & 2.73   & 3.55   & 3.60   & 2.93\\
\%    & -100   & -100   & -50    & 54     & 12     & -6.8   & 22     & 23     & - \\
\hline
\end{tabular}
\end{center}
\end{table}

\begin{table}
\begin{center}
\caption{\label{table4} Ground-state energy of ions from the Li isoelectronic series obtained by GCM-DFT from the LDA and the respective deviation (\%), compared to near-exact reference data \cite{libenchmarkGS}. In the first set of data we use the same mesh already employed for the He ground state, in the second set we used the scaled He mesh of  Eq.(\ref{HeLigsLaw1}): $\alpha = \{3.5;3.9;4.3;4.7;5.1\}$ , as described in the main text, and in the third set we employed the empirical mesh $\alpha = \{3.47;3.73;3.99;4.25;4.51\}$. For comparison, the LDA prediction for Li is $-E_0^{LDA}=7.3440 a.u.$, which deviates from the exact result by $-1.79\%$.}
\begin{tabular}{l|ccccc}
\hline
$^2S$     & $-E^{GCM-DFT}_0$ & $-E^{GCM-DFT}_0$ & $-E^{GCM-DFT}_0$ & $-E^{exact}_0$\cite{libenchmarkGS} \\
& He mesh & Eq.~(\ref{HeLigsLaw1})  & emp. & \\ \hline
$Li$      &  7.3179 (-2.14\%)  & 7.4282 (-0.667\%)  & 7.4742 (-0.0522\%) & 7.4781 \\
$Be^{1+}$ &  14.231 (-0.656\%) & 14.269 (-0.391\%)  & 14.220 (-0.733\%)  & 14.325\\
$B^{2+}$  &  23.146 (-1.19\%)  & 23.367 (-0.248\%)  & 23.335 (-0.384\%)  & 23.425\\
$C^{3+}$  &  34.699 (-0.221\%) & 34.749 (-0.0776\%) & 34.681 (-0.273\%)  & 34.776\\
\hline
\end{tabular}
\end{center}
\end{table}

\begin{table}
\scriptsize
\begin{center}
\caption{\label{table5} First two lines: $2^2S\to3^2S$ excitation energy of the
Li atom, corresponding to the $1s^{2}2s^{1} \to 1s^{2}3s^{1}$ single-particle
transition, obtained from KS, $\Delta$SCF, TDDFT/ALDA and various GCM-DFT
approaches, and its percentage error relative to the near-exact value \cite{libenchmarkES}. Second
set of two lines: Ground-state energy of the Li atom and its percentage error.
Third set of two lines: Energy of the $3^2S$ excited state and its percentage
error. In the column labelled KS we report, instead of total ground-state
and excited-state energies, the KS single-particle energies of the 2s and 3s
state. The ground-state energy in the columns labelled GCM-DFT*, representing
the {\em add-on} use of GCM-DFT is, by definition of the {\em add-on} procedure,
the exact one, while in the columns labelled GCM-DFT, representing the
{\em stand-alone} use of GCM-DFT, it was taken from the GCM-DFT values of
Table~\ref{table4}.}
\begin{tabular}{cccccccccccc}
\hline
 Li & KS & $\Delta$SCF & TDDFT & GCM-DFT & GCM-DFT   & GCM-DFT  & GCM-DFT* & GCM-DFT* & GCM-DFT* & exact \\
 & LDA & LDA & ALDA  & LDA/He  & LDA/scal. & LDA/emp. & LDA/He   & LDA/scal & LDA/emp  & \\ \hline
$2^2S\to3^2S$ & 0.1156 & 0.1199 & 0.1140 & 0.2670 & 0.08080 & 0.1240 & 0.4272 & 0.1307 & 0.1279 & 0.1240 \\
\%          & -6.77 & -3.31  & -8.06  & 115    & -34.8   & 0.00   & 245    & 5.40   & 3.15   & -    \\ \hline
$-E(2^2S)$  & (0.11628) & 7.3440 & 7.3439   & 7.3179  & 7.4282    & 7.4742      & 7.4781 & 7.4781 & 7.4781 & - \\
\%          &  - & -1.79 & -1.79   & -2.14 & -0.667  & -0.0522   & -         & -          & -         & \\\hline
$-E(3^2S)$  & (0.00066) & 7.2241 & 7.2299& 7.0509  & 7.3474    & 7.3502      & 7.0509  & 7.3474    & 7.3502   & 7.3539 \\
\%          & - & -1.77 & -1.69   & -4.12 & -0.0884 & -0.0503   & -4.12  & -0.0884 & -0.0503 &  \\
\hline
\end{tabular}
\end{center}
\end{table}

\end{document}